---

# Crystal structure and physical properties of the Sr-vacant spin-orbit-coupling induced Mott insulator $Sr_{2-x}IrO_4$

Jimei Kong[1,3], S. L. Liu[2,3,*], Jie Cheng[2], Haiyun Wang[2,3],

Xing'ao Li[4,*], Z. H. Wang[5,*]

1. School of Electronic Science and Engineering, Nanjing University of Posts and Telecommunications, Nanjing, 210046, People's Republic of China

2. College of Science, Center of Advanced Functional Ceramics, Nanjing University of Posts and Telecommunications, Nanjing, 210046, People's Republic of China

3. Nanjing University (Suzhou) High-Tech Institute, Suzhou, 215123, People's Republic of China

4. Key Laboratory for Organic Electronics & Information Displays (KLOEID), Institute of Advanced Materials (IAM), Nanjing University of Posts and Telecommunications (NUPT), Nanjing 210046, People's Republic of China

5. National Laboratory of Solid State Microstructures, Department of Physics, Nanjing University, Nanjing, 210093, People's Republic of China

*corresponding author: liusl@njupt.edu.cn (S. L. Liu); iamxali@njupt.edu.cn (Xing'ao Li); zhwang@nju.edu.cn (Z. H. Wang)

**ABSTRACT**

A series of polycrystalline samples of $Sr_{2-x}IrO_4$ ($0 \leq x \leq 0.3$) have been synthesized by a solid-state reaction method. The crystal structure of this doped system can be explained on the basis of the extended nature of 5d electrons and strontium vacancies in $Sr_{2-x}IrO_4$. The analysis of the temperature-dependent resistance of these samples reveals the semiconducting feature, in which the three-dimensional variable range hopping (3D-VRH) behavior is observed at temperatures lower than 120K, the Arrhenius type in intermediate temperatures from 140K to 200K, and the two-dimensional (2D) weak-localization at high temperatures from 220K to 300K. Correspondingly, temperature-dependent magnetic properties in the range of $x \leq 0.30$ can be described by the antiferromagnetically ordered spin system.

## 1. INTRODUCTION

Transition metal oxides (TMO) with a layered perovskite structure have always attracted vigorous research in condensed matter during the past 30 years, mainly because they display various unusual physical phenomena [1,2]. The exotic properties have emerged in the 3d TMOs, such as high-Tc cooper-oxide superconductors attributed to the localized 3d electrons with strong Coulomb interaction. In contrast, the 4d and 5d TMOs are believed to be weakly correlated wide band systems characterized with the weak Coulomb interaction U (0.4~2.5eV) [3-5] of electrons and the enhanced d–p hybridization due to the extended nature of the 4d and 5d electron metal-orbitals. However, novel physics would be induced by the strong spin orbital coupling (SOC) (~0.4eV) [6]proportional to $Z^4$ (Z is the atomic number) together with its competition with other interactions such as Coulomb interaction. One typical phenomenon verified by theoretical calculations and experimental observations is the SOC-induced $J_{eff}$=1/2 Mott insulator in Ir oxides [1,7-8].

Among all the iridates studied, the layered $Sr_2IrO_4$ system is the most well-known relativistic SOC related Mott insulator [7,8]. In essence, strong crystal field splits the 5d band states with $O_h$ symmetry into $t_{2g}$ and $e_g$ orbital states, and the $t_{2g}$ states split into effective total angular momentum $J_{eff}$=1/2 doublet and $J_{eff}$=3/2 quartet bands via the strong SOC [7]. The $Ir^{4+}$ ( $5d^5$ ) ions provide four electrons to fill the $J_{eff}$=3/2 bands, plus one electron to partially fill the $J_{eff}$=1/2 bands, the latter having higher energy. Thus, the original idea is that these compounds should behave as simple metals according to the independent electron picture. Nevertheless, the $J_{eff}$=1/2 band is so narrow that even a reduced Coulomb interaction could open a small gap supporting the insulating state.

The layered $Sr_2IrO_4$ is not only isoelectronic but also isostructural with a crystal structure similar to that of $La_2CuO_4$ [9,10]. A unique and important structural feature of $Sr_2IrO_4$ is the reduced tetragonal structure with space-group $I4_1/acd$ due to the rotation of the $IrO_6$ octahedra about the c axis by ~11°, resulting in a larger unit cell, four times larger than the undistorted cell [10-13]. The most interesting theoretical proposal is the possibility that the unconventional superconductivity could be induced by both the electron and the hole doping in these systems [2,9,14]. For instance, Wang et al. [2] predicted that the doped $Sr_2IrO_4$ should demonstrate unconventional superconductivity after the breaking of its long-range ordered magnetic state. Searching for unconventional superconductivity in these SOC-induced Mott insulators is of great importance, which may open a new area of superconductivity. Many experiments have been taken to search for the possible unconventional superconductivity in the electron-doped $Sr_2IrO_4$. It is reported that a robust metallic state can be found by $La^{3+}$ doped $Sr_{2-x}La_xIrO_4$ single crystals [1], while no metallic behavior is observed in the doped polycrystalline samples [15,16,17]. The metal-insulator transition revealed by resistivity measurements also occurs in $Sr_2IrO_4$ single crystal with dilute oxygen vacancies or in $Ba_2IrO_4$ under pressure at 13.8 GPa [18,19]. Besides, the correction between the electrical and the magnetic properties is found in the Gd-doped $Sr_2IrO_4$ single crystals [20]. However, there are few search for superconductivity by the hole-doping in $Sr_2IrO_4$ system. Hole doping is performed by Rh substitution for Ir in $Sr_2IrO_4$, where a nearly-metallic state is realized [17]. In addition, the insulating state persists in the hole doped $Sr_2IrO_4$ thin films by ionic liquid gating [21]. Anyway, to the best of our knowledge, the unconventional superconductivity has not been found yet in the carrier-doped $Sr_2IrO_4$ system, which is possibly due to that, we believe, the doping level is far lower than that of the theoretical prediction. Moreover, deeply hole-doped $Sr_2IrO_4$ is more promising to achieve a higher transition temperature according to the prediction [22], while less attempts have been made in this area. Hence, more detailed studies are needed to explore the potential superconductivity especially through hole doping.

Here, we perform hole doping into $Sr_{2-x}IrO_4$ with $0\leq x\leq0.3$ by strontium vacancies. The samples will be studied by x-ray diffraction (XRD), the energy dispersive spectrdmeter (EDS), and the Raman scattering together with electrical and magnetic measurements.

## 2. Experimental methods

Polycrystalline samples of $Sr_{2-x}IrO_4$ (x =0.00, 0.05, 0.1, 0.15, 0.20, 0.25, 0.30) are synthesized through the conventional solid-state reaction method as mentioned elsewhere [23]. The initial materials are $IrO_2$ (99.99%) mixed with $SrCO_3$ (99.95%) according to the exact ratio. The weighed materials are sufficiently grinded for 10 hours and pressed into thin pellets(~2mm). They are heated towards 1400K at a rate of 2K/min and kept at this temperature for 40 hours. Then the samples are slowly cooled down to room temperature. The obtained samples are re-grinded for 5 hours to avoid the uniting of crystalline grain during the calcining process. Finally, they are heated again towards 1450K for 40 hours and cooled down to room temperature naturally. The polycrystallines in the range of x=0.00-0.30 are checked by powder x-ray diffraction(XRD) analysis performed on a PANalytical x-ray diffractometer with Cu Kα1 radiation at room temperature. The elemental distribution of the samples is tested by the energy dispersive x-ray spectroscope (EDS). Lattice parameters and Ir-O bond length are derived by Rietveld refinement. Electronic Raman scattering is performed over the doped samples at T=150K. The temperature dependence of the electrical resistivity is measured with a Quantum Design physical property measurement system (PPMS). The magnetic behavior is determined by a Quantum Design magnetic property measurement system (MPMS) employing both zero-field-cooling (ZFC) and field-cooling(FC) methods, with temperature ranging from 10K to 300K at magnetic field 1T.

## 3. RESULTS AND DISCUSSION
### 3.1 Structure characterization

The synthesized polycrystalline samples are analyzed by EDS, XRD and Raman scattering. The typical EDS spectrum in Fig.1 shows that the sample $Sr_{1.9}IrO_4$ consists of Sr, Ir, and O. The comparison between the nominal concentrations of Sr and the real one found from EDS is presented in Fig.2. It can be seen clearly that the data are readily indexed to the stoichiometry.

Displayed in Fig.3(a) are the X-ray powder diffraction patterns of the polycrystalline samples of $Sr_{2-x}IrO_4$ (x=0.00, 0.10, 0.20, 0.30). All the diffraction peaks observed are in good agreement with space group $I4_1/acd$ having $IrO_6$ octahedron rotated around c-axis, rather than tetragonal cells (space group $I4/mmm$) [24]. The rotation angle are calculated from the Rietveld refinements to be 10.49°, 10.41°, 9.46°, 9.15°, 8.71°, 8.10°, 7.49° for x=0, 0.05, 0.10, 0.15, 0.20, 0.25, 0.30, respectively. No reflections of impurity phases are identified from any of the XRD pattern, indicating that the samples are purely synthesized. It is notable that two peaks at 67°and 68° turn into a single one with the increasing of doping intensity, which indicates that the lattice distortions may take place here. Fig.3(b) shows the results of the Rietveld refinement with x=0.10. Shown in Fig.3(c) are the lattice parameters for different doping concentrations by detailed Rietveld refinements of the XRD. The a-axis is lengthened with increasing x, while the c-axis is shortened. The vacancies in the apical strontium can explain the reduction of c-axis. The elongation of a-axis with increasing x is resulted from the reduction of the rotation of the octahedra, and a similar phenomenon is also reported in La doped $Sr_{2-x}La_xIrO_4$ [15]. Notably, the a-axis reflects the Ir-O bond length

A (the in-plane Ir-$O_1$). Table.1 gives the Ir-O bond length A (the in-plane Ir-$O_1$) and C (the apical Ir-$O_2$). Here, the reduction of the c-axis is in contradiction with the extension of apical Ir-O bond length. Similar contradiction is also found in the La doped samples [15], where the a-axis increases with the doping level while the in-plane Ir-O bonds decreasing. It seems that this contradiction is a common feature in this system.

It is found that the variation trends of both A and C are in consistence with the Raman analysis results, shown in Fig.4 respectively. $Sr_2IrO_4$ belongs to the tetragonal space group $I4_1/acd$. Thirty-two Raman active modes are expected according to the factor group analysis [25,26], $\Gamma_{Raman}=4A_{1g}+7B_{1g}+5B_{2g}+16E_g$. The ions of $Sr_2IrO_4$ contribute to all symmetries with the exception of Ir, which does not contribute to the $A_{1g}$ symmetry. The $K_2NiF_4$ ($I4/mmm$) structure yields only four models, $\Gamma_{Raman}=2A_{1g}+2E_g$. Raman scattering is conducted by A.Glamazda et al. [27] to explore the effects of hole doping in single crystals of $Sr_2Ir_{1-x}Ru_xO_4$. It is found that there are no substantial Ru doping effects on the high-frequency phonon modes, so the low-energy phonon modes are discussed instead. Three main peaks at 188, 278, and 392 $cm^{-1}$ correspond respectively to the stretching vibrations of the Sr atoms, bending motions of the Ir-O-Ir bonds and displacements of the in-plane oxygen atoms. However, the evolution has not been discussed of these modes with the doping level. Here, the Raman data are measured in the frequency ranging from 300 to 800 $cm^{-1}$ depicted in Fig.4 and the connection between the lattice parameter and Ir-O bond length is studied. The $A_{1g}$ (560 $cm^{-1}$) and $B_{1g}$ modes (327, 690, and 748 $cm^{-1}$) in Fig.3 can be assigned to modes of oxygen [28,29]. In addition, there are three modes around 370 $cm^{-1}$ so that we can not figure out which mode it belongs to. In particular, the modes of apical oxygen showing up in the frequency range of 500-800 $cm^{-1}$ move towards the lower frequency obviously, which indicates the increasing of Ir-$O_2$ bond length. This is in good agreement with Rietveld analysis of the XRD data.

### 3.2 Resistivity measurements

Fig.5(a) shows the temperature dependence of the electrical resistivity of polycrystalline samples of $Sr_{2-x}IrO_4$ (x =0.05, 0.10, 0.20, 0.25, 0.30). There are no metallic behavior (d$\rho$/dT>0) observed in the entire temperature range from 2K to 300K. All of the samples display the semiconducting temperature dependence. Remarkably, the resistance of $Sr_{2-x}IrO_4$ is systematically decreased with the increasing of defect sites, which suggests that the mobile carriers are exactly introduced into the phases. The insulating state is the consequence of localization due to the disorder on the Sr site and likely the Ir sites feel the effect of this disorder, which is derived from the random potential scattering contributed by a large number of defects or disorders attributed to the vacancy of strontium. However, we have not found any indication of transition from insulator to metal in the strontium vacant $Sr_{2-x}IrO_4$ as we initially expected.

In the insulating state, the temperature dependence of resistivity is analyzed to elucidate the conductivity mechanism of $Sr_{2-x}IrO_4$. In the high temperature region, the resistance data are fitted according to the formula, $\rho(T)=\rho_0\exp(-\alpha T)$, and ln $\rho$ versus $T$ is a straight line over 220K–300K. This behavior has also been reported in $Sr_2IrO_4$ and $Ga_xSr_{2-x}IrO_4$, which is the signature of a 2D weak-localization [30]. In other temperature range, the resistivity data can be described by the following formula:

$$\rho(T) = \rho_0 \exp(\frac{T_0}{T})^{\frac{1}{n}},\qquad(1)$$

where $T_0$ is the localization temperature and n (=1-4) is an integer depending on the conduction mechanism related to the dimensionality of the system[31,32]. When n=1, Eq.(1) indicates the data behave in accordance with the Arrhenius law for thermally activated hopping. When n=2-4, Eq. (1) indicates the variable range hopping(VRH) mechanism [33], with n=2 giving the Efros Shklovskii type VRH scheme particularly [34]. As mentioned by H. Okabe et al [19], carrier conduction in $BaIrO_4$ is dominated by the two-dimensional (2D) VRH behavior, which suggests that the carriers are localized by the disorder in the crystal. All of the models with n=1-4 in Eq. (1) are tested by fitting the data numerically. In the intermediate temperature region from 140K to 200K, Arrhenius type is valid. We mainly focus on the low temperature region depicted in Fig.5(b). The straight lines of the plots of each sample ensure that data are better consistent with the case of n=4, which is known as the three-dimensional variable range hopping (3D-VRH) within the temperature range 2-120K [35]. Our results are consistent with that of $Gd_xSr_{2-x}IrO_4$ [20] and $Sr_{2-x}La_xIrO_4$ [15], while different results are also reported in this temperature region. Carriers can also be doped into $Sr_2IrO_4$ through the electric field effect with ionic liquid gating to change the band filling [21], where gate voltage $V_g$>0(<0) corresponds to n-type (p-type) carriers. The advantage of this approach is that it avoids inducing the chemical disorder. In the case of hole doping (Vg<0), a persistent insulating state is observed, which is consistent with our result. However, it is contradictory that the resistance in the low temperature region follows the 2D-VRH model, which has also been reported by I.N.Bhatti [36]. It seems that this problem is in highly debates.

The fitted localization temperature $T_0$ in Eq. (1) for 3D-VRH systems is inversely proportional to the square of the localization length of the wave function ($\zeta$) and density of states (DOS) at the Fermi level [D(E$_f$)], i.e., $T_0 = 1/k_B D(E_f) \zeta^2$, where $k_B$ is the Boltzmann constant. The values of $T_0$ shown in Table.1 are obtained from the slope of the straight lines in Fig.5(b), which decreases with increasing x concentration approximately. It is discussed that the resistance can be fitted in three different temperature ranges by using two-dimensional VRH mechanism in $Sr_2IrO_4$ [36]. The $D(E_f)$ is calculated from the electronic coefficient of specific heat, which is very low and can be assumed constant. Thus, the value of $T_0$ is consumed merely related with $\zeta$, which is of the order of the in-plane lattice parameter. Our results show that this parameter increases with the doping level, which implies that $\zeta$ also increases with it and therefore $T_0$ decreases. On the other hand, it is reported that the Gd-doped $Sr_2IrO_4$ is a three-dimensional system at low temperatures and the decrease of $T_0$ is the major effect of x doping to increase the DOS at the Fermi level [20]. More experiments such as specific heat are needed to figure out which factor has the major influence on the change of $T_0$.

### 3.3 Magnetic properties

For x=0, $Sr_2IrO_4$ exhibits a clear magnetic transition around $T_C$=240K, which has been reported in both single crystal [37] and polycrystalline samples [38], and such transition also appears in our Sr-vacant samples. A weak ferromagnetic behavior can also be observed in the field cooling (FC) curve of $Sr_2IrO_4$ [15]. It is depicted in Fig. 6(a) that the temperature dependence of dc magnetic measurements under FC (applied field of 1T) and zero field cooling (ZFC; no applied field). A discrepancy between the FC and ZFC curves is observed around 180K for $Sr_{2-x}IrO_4$, and it can be inferred that antiferromagnetic coupling exists in this system. A similar phenomenon has also been reported in $Sr_2Ir_{0.9}Mn_{0.1}O_4$ [38]. This scenario could be understood as follows. The weak ferromagnetism can be attributed to the rotation of the $IrO_6$ octahedra, which changes the behavior in the antiferromagnetical spin system due to the canting moments. However, the antiferromagnetic properties become more obvious with the increasing of doping intensity because of the less rotation of the $IrO_6$ octahedra. The study of $S_2IrO_4$ will further our understanding of the delicate connection between structural distortions and magnetism. A significant feature of strontium-vacant samples is that the values of magnetization decrease with x increasing, which is similar to that of the tensile strain in $Sr_2IrO_4$ thin films [39]. The tensile strain is expected to result in not only the increased electronic bandwidth but also the enhanced electronic correlation. In general, the enhancement of the electronic correlation is in favor of the antiferromagnetic ordering in most Mott insulators, hence the weak ferromagnetism reduces in $Sr_2IrO_4$ system. In addition, the Sr vacancy would probably promote the conservation of equivalent amount of $Ir^{4+}$ into $Ir^{5+}$ of lower spin state, which means decreasing of magnetic moment.

To analyze the susceptibilities, we fit the magnetic data with the Curie-weiss behavior,

$$\chi = \chi_0 + \frac{C}{T - \theta_W} \quad , \qquad (2)$$

where $\chi_0$ is the temperature independence of susceptibility, C is the Curie constant, and $\theta_w$ the Curie-weiss temperature. We then use $\chi_0$ to obtain $\Delta\chi = C/(T-\theta_w)$ and plot the data in terms of $\Delta\chi^{-1}$ vs T, as shown in Fig. 6(b). The undoped $Sr_2IrO_4$ is well known as a weak ferromagnet with a Curie temperature at about 240K. The values of $\theta_w$ are figured to be 205, 222 and 212 K with the corresponding values of x=0.10, 0.20, 0.30. Although the value of $\theta_w$ does not show monotonic behavior, it reduces compared with the undoped sample. Since $\theta_w$ is supposed to measure the strength of the magnetic interaction, the changes of the $\theta_w$ values may be ascribed to the suppression of magnetic ordering. Similar suppression is also reported in $Sr_2Ir_{1-x}Rh_xO_4$ [4].

## 4. Conclusions

In summary, a series of polycrystalline samples of $Sr_{2-x}IrO_4$ have been successfully grown by the solid-state reaction method. The structure and the electrical and magnetic properties of such a system are modified by the strontium vacancies. According to the analysis of structural refinements, the lattice parameters and Ir-O bond length are both changed due to the vacancies in the apical strontium. It can be seen from the temperature-dependent resistance that all the samples display a semiconducting feature, which is in accordance with the 3D-VRH behavior in the low temperature region. Correspondingly, the temperature-dependent magnetization reveals the magnetic anomalies around 180K. Meanwhile, the curves exhibit antiferromagnetic properties under ZFC and FC conditions. Further exploration of $Sr_2IrO_4$ will be more valuable in the future.

## Acknowledgements

We acknowledge the financial support from the Natural Science Foundation for Colleges and Universities in Jiangsu Province (13KJB140012), the National Natural Science Foundation of China (51172110, 11405089), the Natural Science Foundation of Jiangsu Province (BK20130376 , BK20130855), and the Nanotechnology Foundation of Suzhou Bureau of Science and Technology (ZXG201444).

**Explanatory text**

**Fig.1** EDS spectra of $Sr_{1.9}IrO_4$

**Fig.2** The comparison between nominal concentrations of Sr with real one found from EDS of $Sr_{2-x}IrO_4$ (x=0-0.30)

**Fig.3**(a) X-ray powder diffraction patterns of $Sr_{2-x}IrO_4$. (b) X-ray powder diffraction patterns of $Sr_{1.9}IrO_4$ determined by Rietveld refinements. (c) shows the unit-cell parameter a-axis and c-axis of $Sr_{2-x}IrO_4$ at different doping concentrations.

**Fig.4** shows the Raman spectra of $Sr_{2-x}IrO_4$(x=0.10, 0.20, 0.30) measured at 150K.

**Fig.5**(a) Temperature-dependent electrical resistance $\rho$ of $Sr_{2-x}IrO_4$ (x=0.05,0.10,0.20,0.25,0.30) polycrystalline samples. (b) Semi-logarithmic resistance vs. the temperature at minus one-quarter power for $Sr_{2-x}IrO_4$ polycrystalline samples.

**Fig.6**(a) Temperature-dependent magnetization $\chi$ of $Sr_{2-x}IrO_4$(x=0.10, 0.20, 0.30) measured under field cooling (FC) and zero field cooling (ZFC) conditions. (b) displays a Curie-Weiss fit of inverse susceptibility $\Box\chi^{-1}$ (T) for x=0.10, 0.20, 0.30 at high T region.

**Table.1** Units cell parameters, equatorial(A) and apical(C) Ir-O bond lengths in the $IrO_6$ octahedron of $Sr_{2-x}IrO_4$ were calculated from the results of Rietveld refinements.

|  | x=0.00 | x=0.05 | x=0.10 | x=0.15 | x=0.20 | x=0.25 | x=0.30 |
|---|---|---|---|---|---|---|---|
| a (Å) | 5.49541 | 5.49543 | 5.49549 | 5.49587 | 5.4959 | 5.49598 | 5.4962 |
| c (Å) | 25.7946 | 25.7939 | 25.7934 | 25.7931 | 25.7929 | 25.7913 | 25.7885 |
| A (Å) | 2.1507 | 2.1510 | 2.1511 | 2.15117 | 2.15121 | 2.15127 | 2.1513 |
| C (Å) | 1.9757 | 1.9761 | 1.9759 | 1.9765 | 1.9780 | 1.9789 | 1.9793 |

**Table.2** Values of the fitting localization temperature $T_0$ of $Sr_{2-x}IrO_4$ (x=0.05, 0.10, 0.20, 0.25, 0.30) polycrystalline samples.

|  | x=0.05 | x=0.10 | x=0.20 | x=0.25 | x=0.30 |
|---|---|---|---|---|---|
| $T_0$/K | 7.833 | 7.083 | 6.667 | 7.033 | 6.833 |

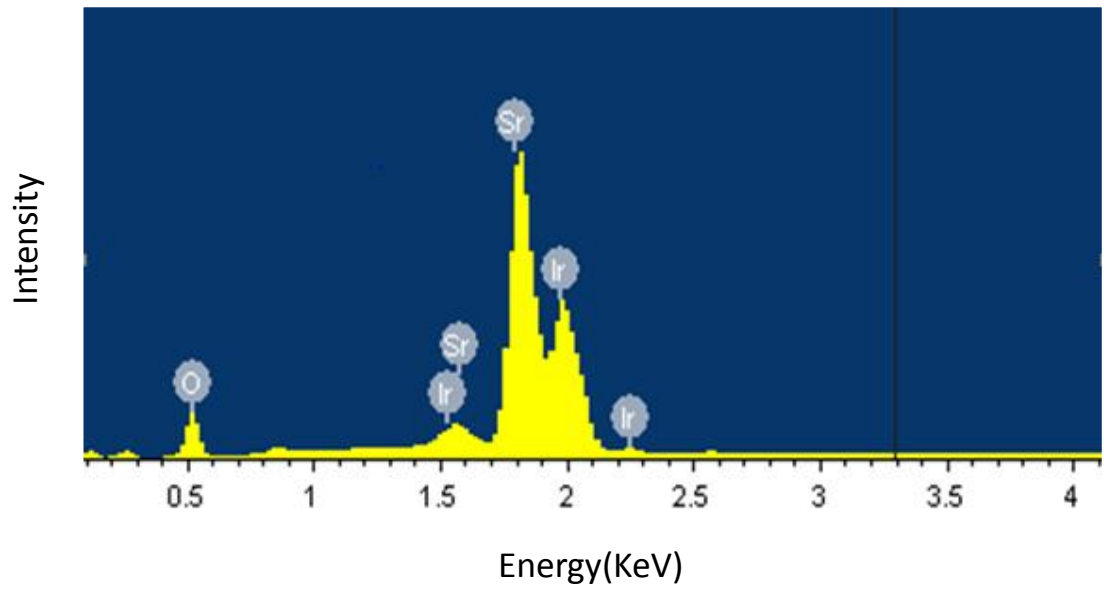

**Fig.1**

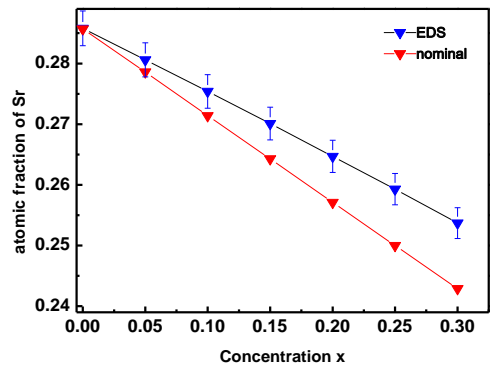

**Fig.2**

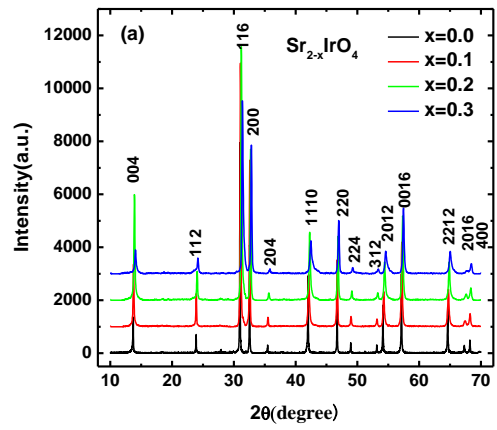

**Fig. 3**(a)

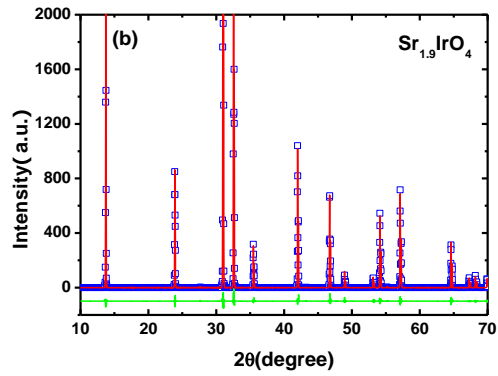

**Fig.3**(b)

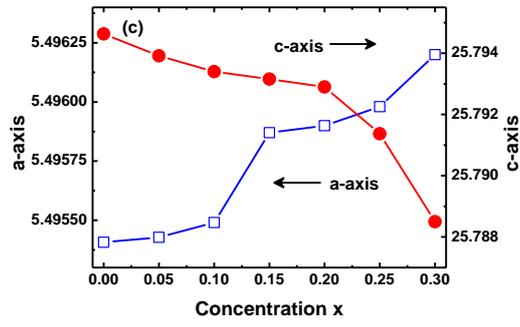

**Fig. 3**(c)

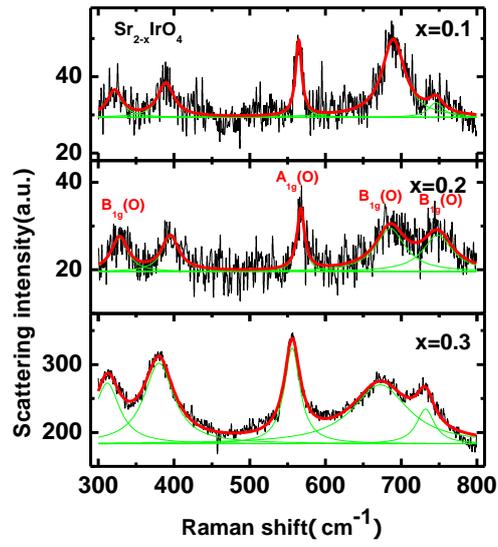

**Fig. 4**

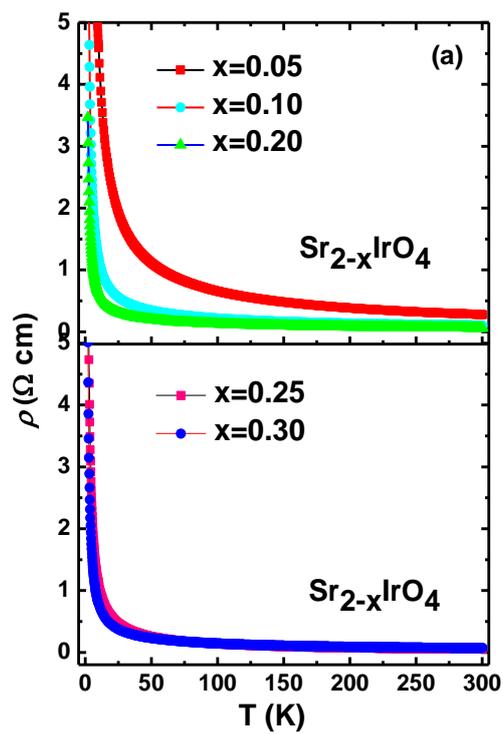

**Fig. 5**(a)

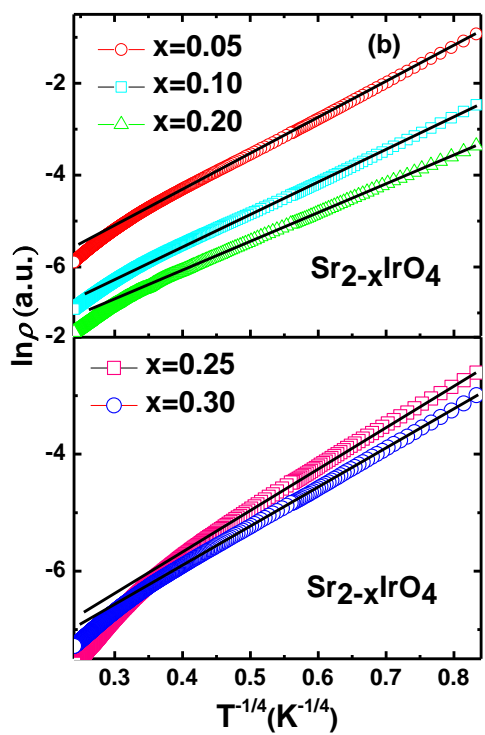

**Fig. 5**(b)

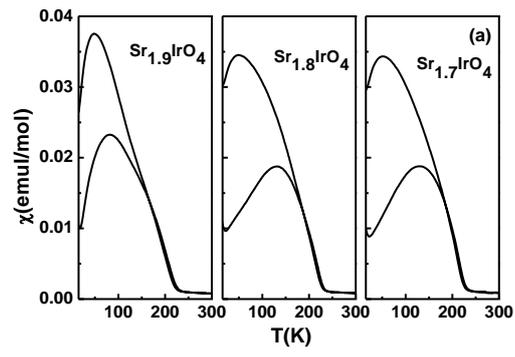

**Fig. 6**(a)

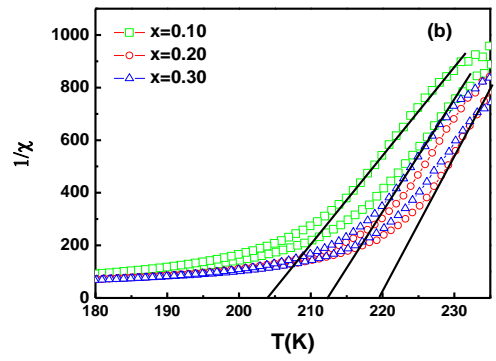

**Fig. 6**(b)